 \documentclass[showpacs,twocolumn,preprintnumbers,amsmath,amssymb]{revtex4}
 \usepackage{graphicx}
 \usepackage{dcolumn}
 \usepackage{bm}
\usepackage{epsfig}
 \begin{document}

 \preprint{}

 \title{The branch processes of vortex filaments and Hopf Invariant Constraint on Scroll Wave}

\author {Tao Zhu}\thanks{Email : zhut05@lzu.cn.}
\author{Ji-Rong Ren}\thanks{Corresponding author. Email: renjr@lzu.edu.cn.}
\author {Shu-Fan Mo}\thanks{Email: meshf07@lzu.cn.}

\affiliation{Institute of Theoretical Physics, Lanzhou University,
Lanzhou 730000, P. R. China}

 \date{\today}

 \begin{abstract}
In this paper, by making use of Duan's topological current theory,
the evolution of the vortex filaments in excitable media is
discussed in detail. The vortex filaments are found generating or
annihilating at the limit points and encountering, splitting, or
merging at the bifurcation points of a complex function
$Z(\vec{x},t)$. It is also shown that the Hopf invariant of knotted
scroll wave filaments is preserved in the branch processes
(splitting, merging, or encountering) during the evolution of these
knotted scroll wave filaments. Furthermore, it also revealed that
the ``exclusion principle" in some chemical media is just the
special case of the Hopf invariant constraint, and during the branch
processes the ``exclusion principle" is also protected by topology.
\end{abstract}

 \pacs{02. 10. Kn, 82. 40. Ck, 02. 40. Xx, 03. 65. Vf}


 \maketitle

\section{Introduction and Motivation}
Scroll waves are three-dimensional (3D) extensions of the familiar
spiral waves of excitable media. They have been observed in a
variety of physical, chemical and biological
systems\cite{chem,phys-CO,phys,bio,BZ,car mus}. Recently, scroll
wave have drawn great interest due to its importance in the the
mechanism of some re-entrant cardiac arrhythmias and fibrillation
which is the leading cause of death in the industrialized
world\cite{7,8,9}. The scroll wave rotate about a linelike filaments
called vortex filament, and usually can be defined in terms of a
phase singularity. In three-dimensional excitable media, the vortex
filament is commonly a closed ring, and these vortex filaments can
form linked and knotted rings which contract to compact,
particle-like
bundles\cite{knot1,knot2,knot4,knot5,knot6,topo2,hopfscroll,topo3,topo4,knot3,topo1,topo5}.

Control of scroll wave is a more important and complex problem for
all excitable media. The dynamics of a 3D scroll wave are determined
not only by properties of the excitable media but also by the
geometry and topology of vortex filament\cite{car mus,7,8,9}. This
implies that the control of scroll wave should stronger depend on
our understanding on scroll wave topology. This inspirits us to use
the topological viewpoint to study the scroll wave topology. In
previous
works\cite{topo2,hopfscroll,topo3,topo4,knot3,topo1,topo5,zero,mo},
many authors have made great contributions to this issue and
employed topological arguments to understand scroll wave. Some most
important topological constraints on behaviors of the vortex
filaments have been investigated. These topological rules may have
some important applications in practice. In particular, the
topological constraint on knotted vortex filaments is believed to
relate to topological characteristic numbers of knotted vortex
filament family, such as the winding, the self-linking and the
linking numbers. In Ref.\cite{knot3}, Winfree and Strogatz  have
proposed an "exclusion principle" which governed the scroll wave
knotting and linking through each other in chemical system. This
exclusion principle gives the constraint on the linking numbers and
winding numbers of scroll wave.

Recently, Duan's topological current
theory\cite{topo2,hopfscroll,mo,hopf,topo-cur1,topo-cur2} has been
applied to study the topological properties of spiral waves and
scroll waves. Zhang et al.\cite{topo2} presented a rigorous
topological description of spiral waves and scroll waves. They
derived precise expressions of spiral wave and scroll wave
topological charge density. Based on their work, we study the branch
process of the spiral waves and calculate the knotted invariant for
knotted vortex filaments by using the Duan's topological current
theory, we proposed that the knotted invariant (which is just the
Hopf invariant) may imply a new topological constraint on scroll
wave\cite{hopfscroll,mo}. However, in Ref.\cite{topo2} and in our
previous work\cite{hopfscroll}, the discussions are based on an
important regular condition $D(\frac{\phi}{x})\neq 0$. When this
condition fails, what will happen? Main purpose of this paper is to
detail this problem.

In this paper, by making using Duan's topological theory, firstly,
we will extend our branch theory of spiral wave in 2D to 3D scroll
wave, and study the generating, annihilating, colliding, splitting
and merging of vortex filaments from a topology viewpoint. Secondly,
based on the branch process of vortex filaments, it is showed that
the Hopf invariant of knotted scroll wave is preserved in the branch
process. This is consistent with our proposal that the Hopf
invariant may implies a new constraint on scroll wave. Third, it is
also shown that the ``exclusion principle" in some chemical media is
just the special case of the Hopf invariant constraint, and during
the branch processes the ``exclusion principle" is also protected by
topology.
\section{Topological Structures of vortex filaments}
In order to maintain the continuity of the whole work and make the
background of this paper clear, in this section, we give a brief
review of the topological current theory of vortex filaments. We
chose to work with a general two-variable reaction-diffusion system
whose mathematical description in terms of a nonlinear partial
differential equation. This equation is written as
\begin{eqnarray}
\partial_t u&=&f(u,v)+D_u \nabla^2 u,\nonumber\\
\partial_t v&=&g(u,v)+D_v \nabla^2v,\label{rd}
\end{eqnarray}
where $u$ and $v$ represent the concentrations of the reagents;
$\nabla^2$ is the Laplacian operator in three-dimensional space;
$f(u,v)$ and $g(u,v)$ are the reaction functions. Following the
description in Ref.\cite{topo2,time}, we define a complex function
$Z=\phi^1+i\phi^2$, where $\phi^1=u-u^*$ and $\phi^2=v-v^*$. Here
$u^*$ and $v^*$ are the concentrations of the vortex filaments.

We know that the complex function $Z=\phi^1+i\phi^2$ can be regarded
as the complex representation of a two-dimensional vector field
$\vec{Z}=(\phi^1,\phi^2)$. Let us define the unit vector:
$n^a=\frac{\phi^a}{\|\phi\|} (a=1,2; \|\phi\|^2=\phi^a\phi^a=Z^*Z)$.
It is easy to see that the zeros of $Z$ are just the singularities
of $\vec{n}$. Using this unit vector $\vec{n}$, an ``induced abelian
gauge potential" can be constructed with
\begin{eqnarray}
A_\mu&=&\epsilon_{ab}n^a\partial_\mu n^b,~~~~~~\mu=0,1,2,3;\nonumber\\
\partial_\mu&=&(\partial_0,
\nabla),~~~\partial_0=\partial_t=\frac{\partial}{\partial t},
\end{eqnarray}
the gauge field strength given by this gauge potential is
\begin{eqnarray}
F_{\mu\nu}=\partial_\mu A_\nu-\partial_\nu A_\mu.
\end{eqnarray}
According to
Ref.\cite{topo2,hopfscroll,mo,hopf,topo-cur1,topo-cur2}, the two
dimensional topological tensor current is defined as
\begin{eqnarray}
K^{\mu\nu}=\frac{1}{4\pi}\epsilon^{\mu\nu\lambda\rho}F_{\lambda\rho}=
\frac{1}{2\pi}\epsilon^{\mu\nu\lambda\rho}\partial_\lambda n^a
\partial_\rho n^b.\label{tensor1}
\end{eqnarray}
It is easy to show that the topological tensor current $K^{\mu\nu}$
can be rewritten in a compact form
\begin{eqnarray}
K^{\mu\nu}=\delta^2(\vec{\phi})D^{\mu\nu}(\frac{\phi}{x}),\label{tensor}
\end{eqnarray}
where $D^{\mu\nu}$ is the general Jacobian determinants
\begin{eqnarray}
\epsilon^{ab}D^{\mu\nu}(\frac{\phi}{x})=\epsilon^{\mu\nu\lambda\rho}\partial_\lambda
\phi^a \partial\rho \phi^b.
\end{eqnarray}
Defining the spatial components of $K^{\mu\nu}$ as
\begin{eqnarray}
j^i=K^{0i}=\frac{1}{2\pi}\epsilon^{ijk}\epsilon_{ab}\partial_j n^a
\partial_k n^b,~i,j,k=1,2,3,\label{topcu}
\end{eqnarray}
we have
\begin{equation}
j^i=\delta(\vec{\phi})D^i(\frac{\phi}{x}),
\end{equation}
where
$D^i(\frac{\phi}{x})=\frac{1}{2}\epsilon^{ijk}\epsilon_{ab}\partial_j
\phi^a \partial_k \phi^b$ is the Jacobian vector. This delta
function expression of the topological current $j^i$ tells us it
doesn't vanish only when the vortex filaments exist, i.e.,
\begin{equation}
j^i\left\{
      \begin{array}{ll}
        =0, & \hbox{if and only if $\vec{\phi}\neq 0$;} \\
        \neq 0, & \hbox{if and only if $\vec{\phi}=0$.}
      \end{array}
    \right.
\end{equation}
So the sites of the vortex filaments determine the nonzero solutions
of $j^i$. The implicit function theory shows that under the regular
condition\cite{imp}
\begin{equation}
D^{\mu\nu}(\frac{\phi}{x})\neq 0,\label{recon}
\end{equation}
the general solutions of
\begin{equation}
\phi^1(t,\vec{x})=0,~~\phi^2(t,\vec{x})=0\label{zero}
\end{equation}
can be expressed as
\begin{equation}
x^1=x^1_l(t,s),~~x^2=x^2_l(t,s),~~x^3=x^3_l(t,s),
\end{equation}
which represent the world surface of $N$ moving isolated vortex
filaments with string parameter $s$ ($l=1,2,\cdot,N$). These
singular strings solutions are just the vortex filaments. In delta
function theory\cite{del}, one can prove that in three-dimensional
space,
\begin{equation}
\delta(\vec{\phi})=\sum_{l=1}^{N} \beta_k \int_{L_l}
\frac{\delta^3(\vec{x}-\vec{x}_l(s))}{|D(\frac{\phi}{u})|_{\Sigma_l}}ds,\label{8}
\end{equation}
where
$D(\frac{\phi}{u})=\frac{1}{2}\epsilon^{jk}\epsilon_{mn}\frac{\partial
\phi^m}{\partial u^j} \frac{\partial \phi^n}{\partial u^k}$ and
$\Sigma_l$ is the $l$-th planar element transverse to $L_l$ with
local coordinates $(u^1,u^2)$. The positive integer $\beta_l$ is the
Hopf index of $\phi$-mapping, which means that when $\vec{x}$ covers
the neighborhood of the zero point $\vec{x}_l(s,t)$ once, the vector
field $\vec{\phi}$ covers the corresponding region in $\phi$ space
for $\beta_l$ times. Meanwhile the direction vector of $L_l$ is
given by\cite{topo-cur1,topo-cur2}
\begin{equation}
\frac{dx^i}{ds}|_{\vec{x}_l}=\frac{D^i(\phi/x)}{D(\phi/u)}|_{\vec{x}_l}.\label{9}
\end{equation}
Then considering Eqs.(\ref{8}) and Eqs.(\ref{9}), we obtain the
inner structure of $j^i$,
\begin{eqnarray}
j^i&=&\delta(\vec{\phi})D^i(\frac{\phi}{x})\nonumber\\
&=&\sum_{l=1}^{N} \beta_l \eta_l \int_{L_l} dx^i
\delta^3(\vec{x}-\vec{x}_l),\label{10}
\end{eqnarray}
where $\eta_l=sgn D(\frac{\phi}{u})=\pm 1$ is the Brouwer degree of
$\phi$-mapping, with $\eta_l=1$ corresponding to the vortex filament
and $\eta_l=-1$ corresponding to the antivortex filament. We find
that the topological current $\vec{j}$ is just the charge density
vector $\vec{\rho}$ of the vortex filament in Ref.\cite{topo2}. In
our theory, the topological charge of the vortex filament $L_l$ is
\begin{equation}
Q_l=\int_{\Sigma_l} \vec{j}\cdot d\vec{\sigma}=W_l=\beta_l
\eta_l,\label{11}
\end{equation}
in which $W_l$ is just the winding number of $\vec{\phi}$ around
$L_l$, the above expression reveals distinctly that the topological
charge of vortex filament is not only the winding number, but also
expressed by the Hopf indices and Brouwer degrees. The topological
inner structure showed in Eq.(\ref{11}) is more essential than
usually considered and it will be helpful as a complement of the
current description of scroll wave only by winding number in
topology. This shows that the advantage of our topological
description of the vortex filaments.

In the above, we give an prime introduction of the topological
current theory of scroll wave, which is based on the rigorous
mathematics background that we called Duan's topological current
theory. Here we give some remarks about above results in order.
\emph{(i)}. The definition of the complex function
$Z=\phi^1+i\phi^2$ originates from $u$ and $v$. Therefore, $Z$ is
the function that shows the concentration distribution of the
reagents. The exact expression of $Z$ is determined by the nonlinear
partial differential equation (\ref{rd}). If the exact analytical
solution of Eq.(\ref{rd}) is known, we can directly calculate the
zero points of $Z$ and determine the Hopf indices and Brouwer
degrees, which describe the inner topological structures of the
vortex filaments of scroll wave. \emph{(ii)}. The traditional
theoretical description of scroll wave usually used the phase
singularity method and cannot directly deal with the zero points of
$Z$. The above results show that the topological current theory of
scroll wave provide a available approach to deal with the zero
points. \emph{(iii)}. The regular condition (\ref{recon}) will plays
a essential role in determining the stability of the scroll wave.
This condition also depends on the exact solution of the
Eq.(\ref{rd}). It can be determined when the expression of $Z$ is
known. How this condition determine the stability of the scroll wave
is the tasks of the following sections.

\section{The branch process of vortex filaments at the limited point}
However, from the above discussion we know that the results
mentioned are obtained under the condition
$D^{\mu\nu}(\frac{\phi}{x})\neq 0$. When this condition fails, i.e.,
the Brouwer degrees $\eta_l$ are indefinite, what will happen? In
what follows, we will study the case when
$D^{\mu\nu}(\frac{\phi}{x})=0$. It often happens when the zero of
$\vec{Z}$ includes some branch points, which lead to the bifurcation
of the topological current.

Generally speaking, the evolution of a vortex filament $L_l$ can be
discussed from Eq.(\ref{tensor}). From Eq.(\ref{tensor1}),
considering that $\epsilon^{\mu\nu\lambda\rho}$ is a fully
antisymmetric tensor, we can prove that
\begin{eqnarray}
\partial_\mu K^{\mu\nu}=0,\label{cq2}
\end{eqnarray}
that is
\begin{eqnarray}
\partial_0 j^i+\partial_j K^{ji}=0.
\end{eqnarray}
This is the continuity equation constraints on vortex filaments. In
order to discuss the evolution of these vortex filaments and
simplify our study, we fixed the $x^3=z$ coordinate and take the
$XOY$ plane as the cross section, so the intersection line between
the $L_l$'s evolution surface and the cross section is just the
motion curve of $L_l$. In this case the 2D topological current is
defined as
\begin{eqnarray}
j^3=K^{03}=\delta^2(\phi)D^0(\frac{\phi}{x})
\end{eqnarray}
and
\begin{eqnarray}
K^i=K^{i3}=\delta^2(\phi)D^i(\frac{\phi}{x}), ~~i=1,2.
\end{eqnarray}
It is easy to see that $j^3$ and $K^i$ satisfy the continuity
equation
\begin{eqnarray}
\partial_0 j^3+\partial_i K^{i}=0.\label{cq}
\end{eqnarray}
The velocity of the intersection point of vortex filament and the
cross section is given by
\begin{eqnarray}
\frac{dx^i}{dt}=\frac{D^i(\phi/x)}{D^0(\phi/x)}.\label{ve}
\end{eqnarray}

From Eq.(\ref{ve}) it is obvious that when
$$D^0(\phi/x)=0$$
at the very point $(t^*,\vec{x}^*)$, the velocity
\begin{eqnarray}
\frac{dx^1}{dt}=\left.\frac{D^1(\phi/x)}{D^0(\phi/x)}\right|_{(t^*,\vec{x}^*)},
\frac{dx^2}{dt}=\left.\frac{D^2(\phi/x)}{D^0(\phi/x)}\right|_{(t^*,\vec{x}^*)}
\end{eqnarray}
is not uniquely determined in the neighborhood of $(t^*,\vec{x}^*)$.
This critical point is called the branch point. In Duan's
topological current theory usually there are two kinds of branch
points, namely the limit points and the bifurcation points, each
kind of which corresponds to different cases of branch process.

First, in this section, we only study the case that the zeros of the
complex function $\vec{Z}$ includes some limit points which satisfy
\begin{eqnarray}
\left.D^1(\frac{\phi}{x})\right|_{(t^*,\vec{x}^*)}\neq 0,
~~~\left.D^2(\frac{\phi}{x})\right|_{(t^*,\vec{x}^*)}\neq 0.
\end{eqnarray}
For simplicity, we assume that $D^2(\phi/x)|_{(t^*,\vec{x}^*)}\neq
0$ is always satisfied in our next discussions. When
$D^1(\phi/x)|_{(t^*,\vec{x}^*)}\neq 0$, from Eq.(\ref{ve}) we obtain
\begin{eqnarray}
\frac{dx^1}{dt}=\left.\frac{D^1(\phi/x)}{D^0(\phi/x)}\right|_{(t^*,\vec{x}^*)}
=\infty;
\end{eqnarray}
i.e.,
\begin{eqnarray}
\left.\frac{dt}{dx^1}\right|_{(t^*,\vec{x}^*)}=0.
\end{eqnarray}
Then, the Taylor expansion of of $t=t(x^1,t)$ at the limit point
$(t^*,\vec{x}^*)$ of vortex filament, one can obtain
\begin{eqnarray}
t-t^*=\frac{1}{2}\left.\frac{d^2t}{(dx^1)^2}\right|_{(t^*,\vec{x}^*)}(x^1-x^{1*})^2,\label{lim}
\end{eqnarray}
which is a parabola in $x^1-t$ plane. From Eq.(\ref{lim}) we can
obtain two solutions $x^1_1(t)$ and $x^1_2(t)$, which give two
branch solutions of vortex filament at the limit points. If
\begin{eqnarray}
\left.\frac{d^2t}{(dx^1)^2}\right|_{(t^*,\vec{x}^*)}>0,\label{lim2}
\end{eqnarray}
we have the branch solutions for $t>t^*$ [see Fig. 1(a)]; otherwise,
we have the branch solutions for $t<t^*$[see Fig. 1(b)]. The former
is related to the origin of the vortex filament at the limit points,
and the later is the annihilation of the vortex filament. At the
neighborhood of the limit point, we denote the length scale
$l=\Delta x$. From Eq.(\ref{lim}), one can obtain the approximation
relation
\begin{eqnarray}
l\propto \parallel t-t^*\parallel^{1/2}.
\end{eqnarray}
The growth rate $\gamma=l/\Delta x$ or annihilation rate of vortex
lines is
\begin{eqnarray}
\gamma\propto (t-t^*)^{-1/2}.
\end{eqnarray}

From the continuity equation Eq.(\ref{cq}), we know that the
topological number of the vortex filament is identically conserved.
This means that the total topological number of the final vortex
filaments equals that of the initial vortex filaments. The total
numbers of these two generated or annihilated vortex filaments must
be zero at the limit point; i.e., the two generated or annihilated
vortex filaments have be opposite,
\begin{eqnarray}
\beta_1\eta_1+\beta_2\eta_2=0,
\end{eqnarray}
which shows that $\beta_1=\beta_2$ and $\eta_1=-\eta_2$. One can see
the fact that the Brouwer degree $\eta$ is indefinite at the limit
points implies that it can change discontinuously at limit points.

For a limit point it is required that
$D^1(\phi/x)|_{(t^*,\vec{x}^*)}\neq 0$. As to a bifurcation
point\cite{bif}, it must satisfy a more complex condition. This case
will be discussed in the following section.

\begin{figure}[h]
\begin{center}
\begin{tabular}{p{5cm}p{5cm}}
(a)\psfig{figure=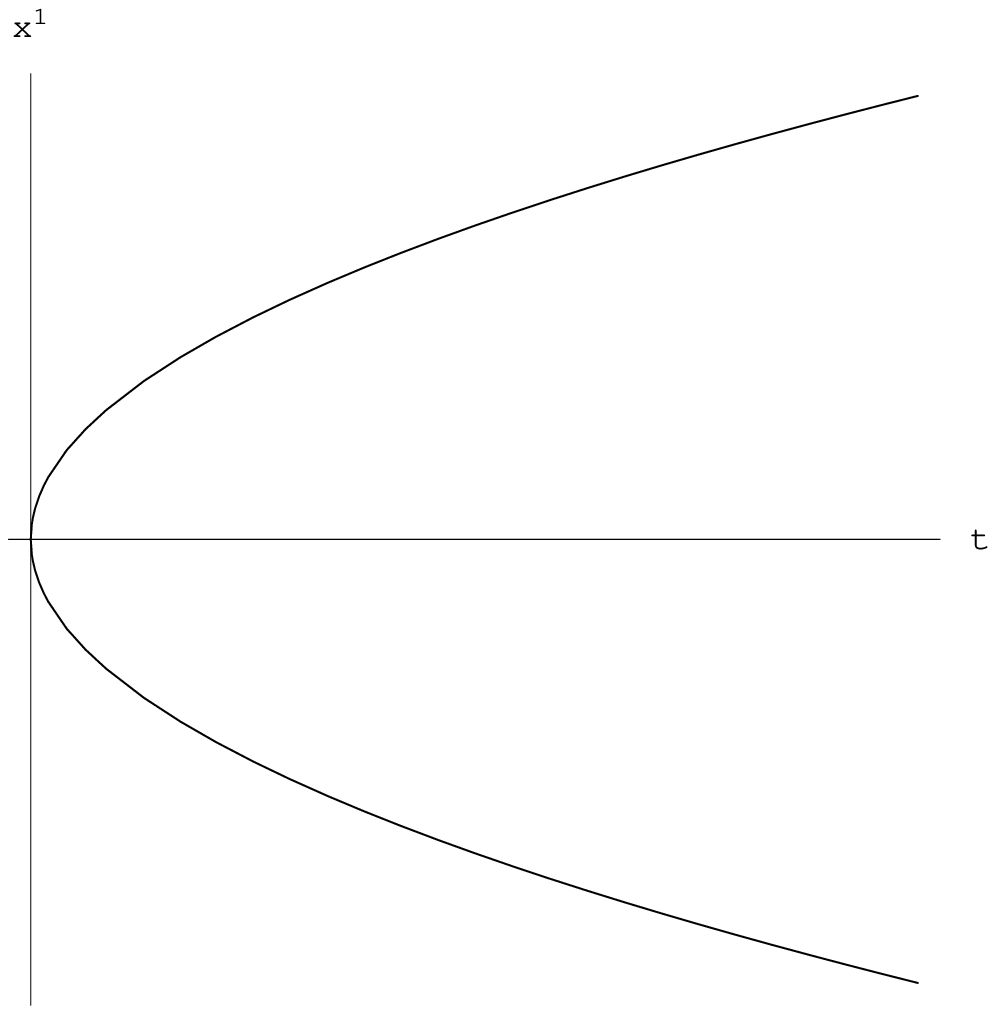,height=5cm,width=5cm}\\
(b)\psfig{figure=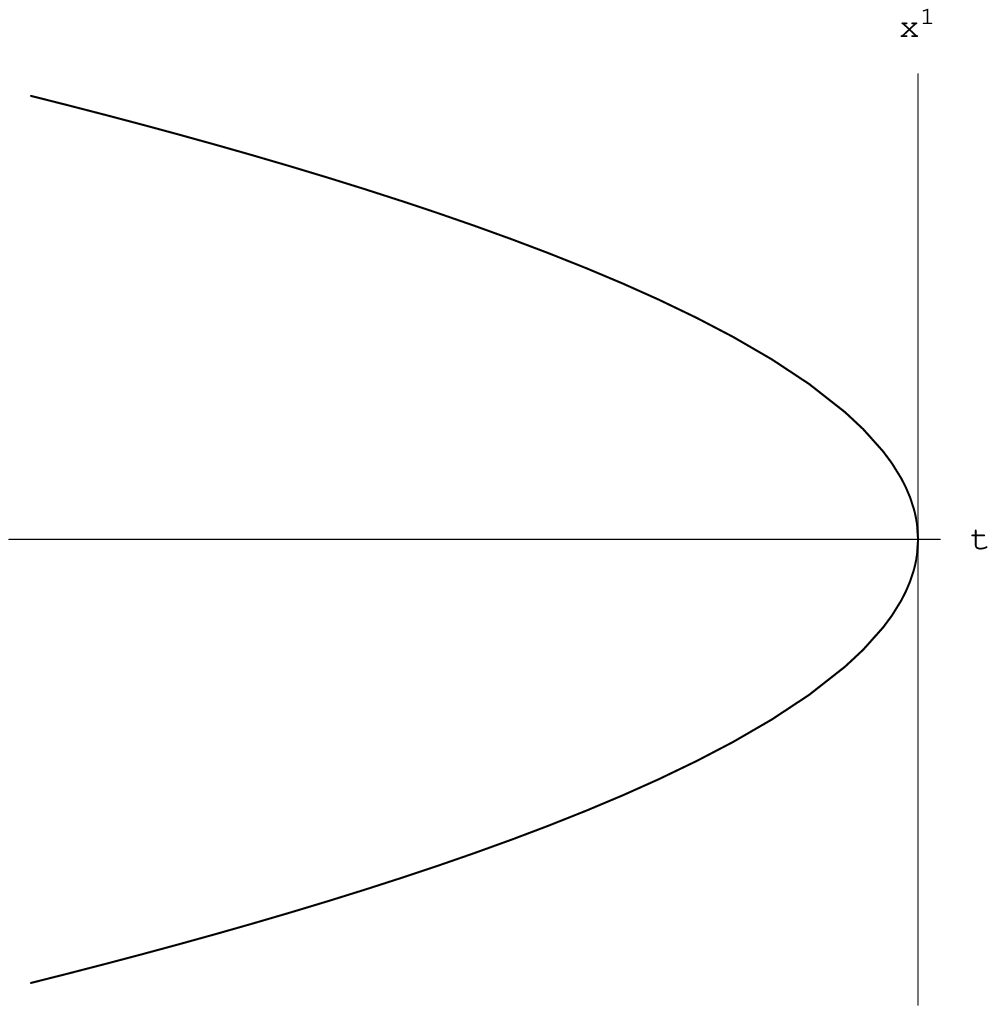,height=5cm,width=5cm}
\end{tabular}
\end{center}
\caption{We fixed the limit point $(x^{1*},t^*)$ at the origin of
$(x^1-t)$ plane. (a) The branch solutions for Eq.(\ref{lim}) when
$d^2t/(dx^1)^2|_{(t^*,\vec{x}^*)}>0$, i.e., a pair of vortex
filaments with opposite charges generate at the limit point, i.e.,
the origin of vortex filaments. (b) The branch solutions for
Eq.(\ref{lim}) when $d^2t/(dx^1)^2|_{(t^*,\vec{x}^*)}<0$, i.e., a
pair of vortex filaments with opposite charges annihilate at the
limit point.} \label{fig1}
\end{figure}

\section{The branch process of vortex filaments at the bifurcation point}
Now let us study the bifurcation of the vortex line at its
bifurcation point where
\begin{eqnarray}
\left.D^0(\frac{\phi}{x})\right|_{(t^*,\vec{x}^*)}=0,
~~\left.D^1(\frac{\phi}{x})\right|_{(t^*,\vec{x}^*)}=0.\label{bifp}
\end{eqnarray}
These two restrictive conditions will lead to an important fact that
the function relationship between $t$ and $x^1$ is not unique in the
neighborhood of the bifurcation point $(t^*,\vec{x}^*)$. The
equation
\begin{eqnarray}
\frac{dx^1}{dt}=\left.\frac{D^1(\phi/x)}{D^0(\phi/x)}\right|_{(t^*,\vec{x}^*)},\label{bif1}
\end{eqnarray}
which, under restraint of Eq.(\ref{bifp}), directly shows that the
direction of the integral curve of Eq. (\ref{bif1}) is indefinite at
the point $(t^*,\vec{x}^*)$. This is why the very point
$(t^*,\vec{x}^*)$ is called a bifurcation point.

Assume that the bifurcation point $(t^*,\vec{x}^*)$ has been found
from Eqs.(\ref{zero}) and (\ref{bifp}). We know that, at the
bifurcation point $(t^*,\vec{x}^*)$, the rank of the Jacobian matrix
$[\partial\phi/\partial x]$ is $1$. In addition, according to the
Duan's topological current theory, the Taylor expansion of the
solution of Eq.(\ref{zero}) in the neighborhood of the  bifurcation
point $(t^*,\vec{x}^*)$ can be expressed as\cite{topo-cur1,mo}
\begin{eqnarray}
A(x^1-x^{*1})^2+2B(x^1-x^{*1})(t-t^*)+C(t-t^*)^2=0,
\end{eqnarray}
which leads to
\begin{eqnarray}
A(\frac{dx^1}{dt})^2+2B(\frac{dx^1}{dt})+C=0\label{b1}
\end{eqnarray}
and
\begin{eqnarray}
C(\frac{dt}{dx^1})^2+2B(\frac{dt}{dx^1})+A=0,\label{b2}
\end{eqnarray}
where $A$, $B$, and $C$ are three constants. The solutions of
Eq.(\ref{b1}) or Eq.(\ref{b2}) give different directions of the
branch curves at the bifurcation point. There are four possible
cases, which will show the physical meanings of the bifurcation
points.
\begin{figure}[h]
\begin{center}
\begin{tabular}{p{5cm}p{5cm}}
\psfig{figure=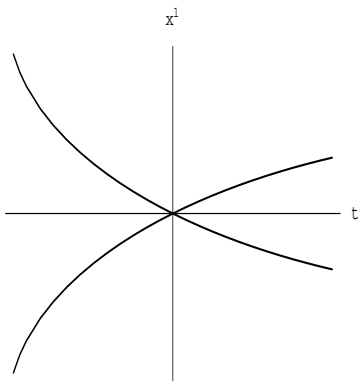,height=5cm,width=5cm}
\end{tabular}
\end{center}
\caption{We fixed the bifurcation point $(x^{1*},t^*)$ at the origin
of $(x^1-t)$ plane. Two vortex filaments meet and then depart at the
bifurcation point.} \label{fig2}
\end{figure}

\emph{Case 1 ($A\neq 0$)}. For $\Delta=4(B^2-AC)>0$ from
Eq.(\ref{b1}) we get two different directions of the velocity field
of vortex filaments
\begin{eqnarray}
\left.\frac{dx^1}{dt}\right|_{(t^*,\vec{x}^*)}=\frac{-B\pm
\sqrt{B^2-AC}}{A},\label{case1}
\end{eqnarray}
Which is shown in Fig.2. It is the intersection of two vortex
filaments with different directions at the bifurcation point, which
means that two vortex filaments meet and then depart from each other
at the bifurcation point.

\emph{Case 2 ($A\neq 0$)}. For $\Delta=4(B^2-AC)=0$ from
Eq.(\ref{b1}) we obtain only one direction of the velocity of vortex
filaments
\begin{equation}
\left.\frac{dx^1}{dt}\right|_{(t^*,\vec{x}^*)}=\frac{-B}{A}\label{case2}
\end{equation}
which includes three important situations. (a)Two vortex filaments
tangentially encounter at the bifurcation point [See Fig.3(a)].
(b)Two vortex filaments merge into one vortex filament at the
bifurcation point [See Fig.3(b)]. (c) One vortex filament splits
into two vortex filaments at the bifurcation point [See Fig.3(c)].

\begin{figure}[h]
\begin{center}
\begin{tabular}{p{5cm}p{5cm}}
(a)\psfig{figure=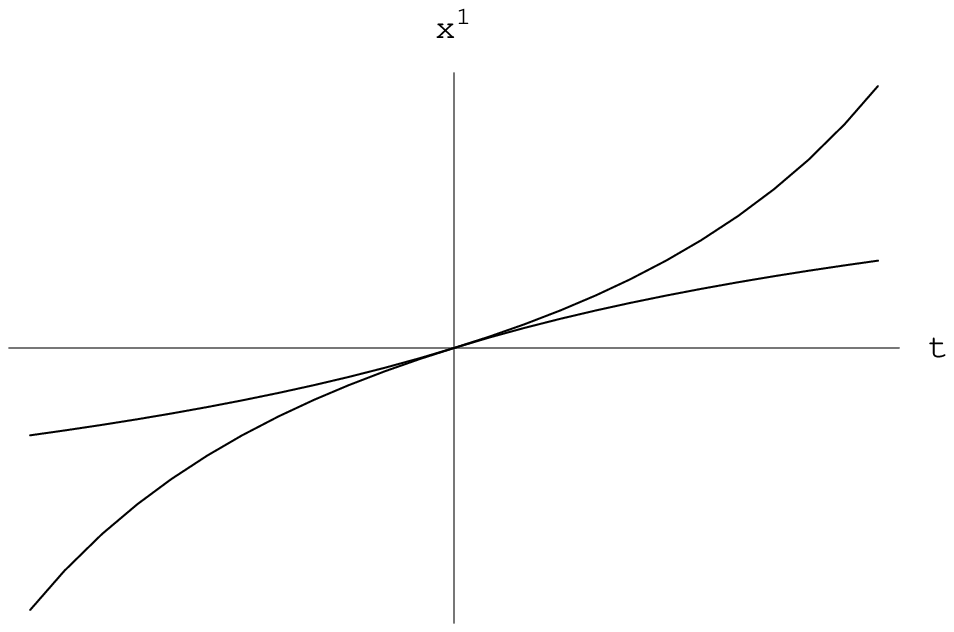,height=5cm,width=5cm}\\
(b)\psfig{figure=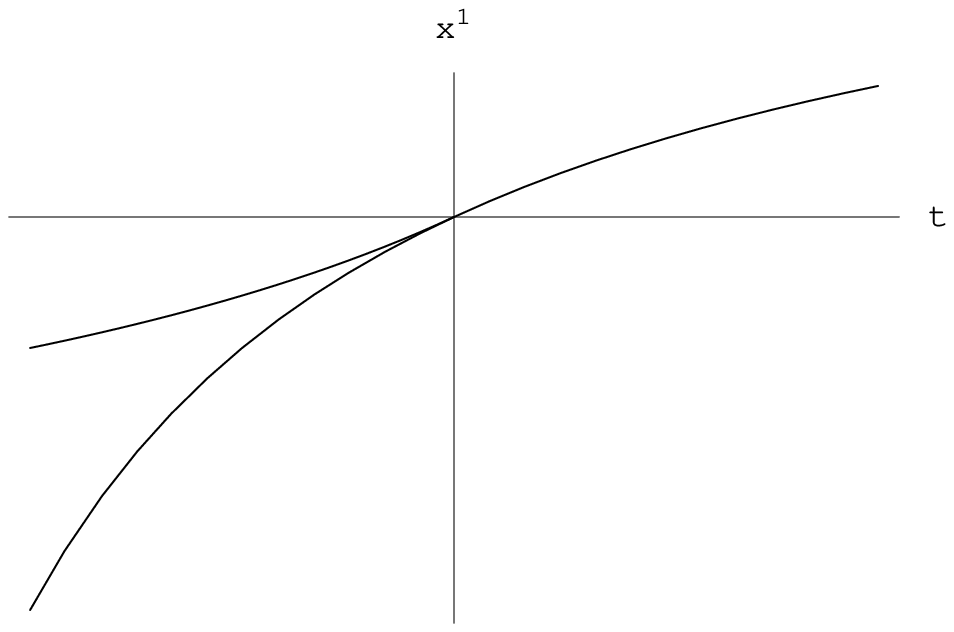,height=5cm,width=5cm}\\
(c)\psfig{figure=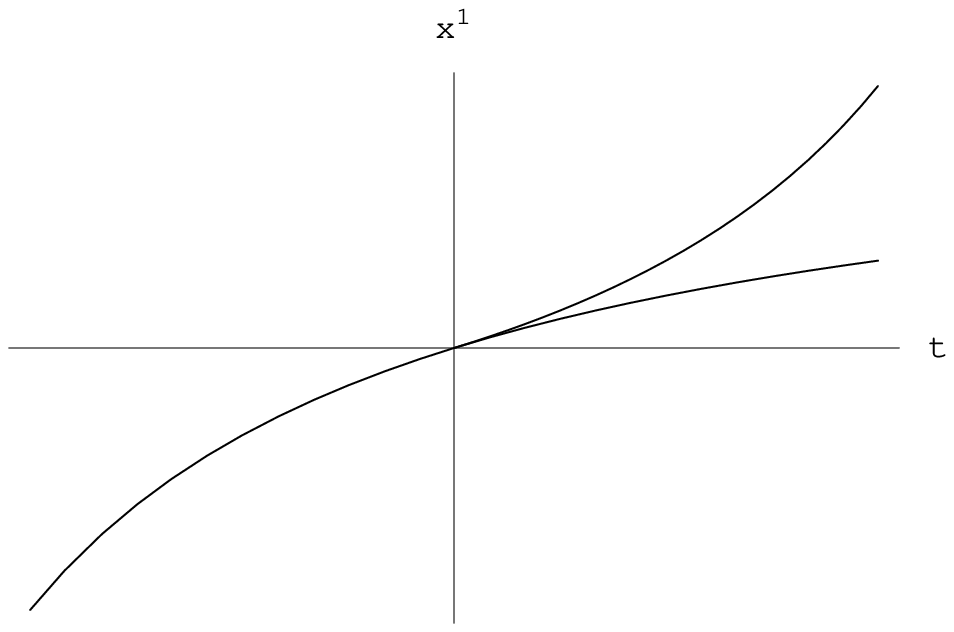,height=5cm,width=5cm}
\end{tabular}
\end{center}
\caption{We fixed the bifurcation point $(x^{1*},t^*)$ at the origin
of $(x^1-t)$ plane. (a) Two vortex filaments tangentially encounter
at the bifurcation point. (b) Two vortex filaments merge into one
vortex filament at the bifurcation point. (c)One vortex filament
splits into two vortex filaments at the bifurcation point.}
\label{fig3}
\end{figure}

\emph{Case 3 ($A=0, C\neq 0$)}. For $\Delta=4(B^2-AC)=0$, we have
\begin{equation}
\left.\frac{dt}{dx^1}\right|_{1,2}=\frac{-B\pm\sqrt{B^2-AC}}{C}=\left\{
                                     \begin{array}{ll}
                                       0, \\
                                       -\frac{2B}{C}.
                                     \end{array}
                                   \right.
\label{case3}
\end{equation}
There are two important cases: (a) One vortex filament splits into
three vortex filaments at the bifurcation point [See Fig.4(a)]. (b)
Three vortex filaments merge into one vortex filament at the
bifurcation point [See Fig.4(b)].

\begin{figure}[h]
\begin{center}
\begin{tabular}{p{5cm}p{5cm}}
(a)\psfig{figure=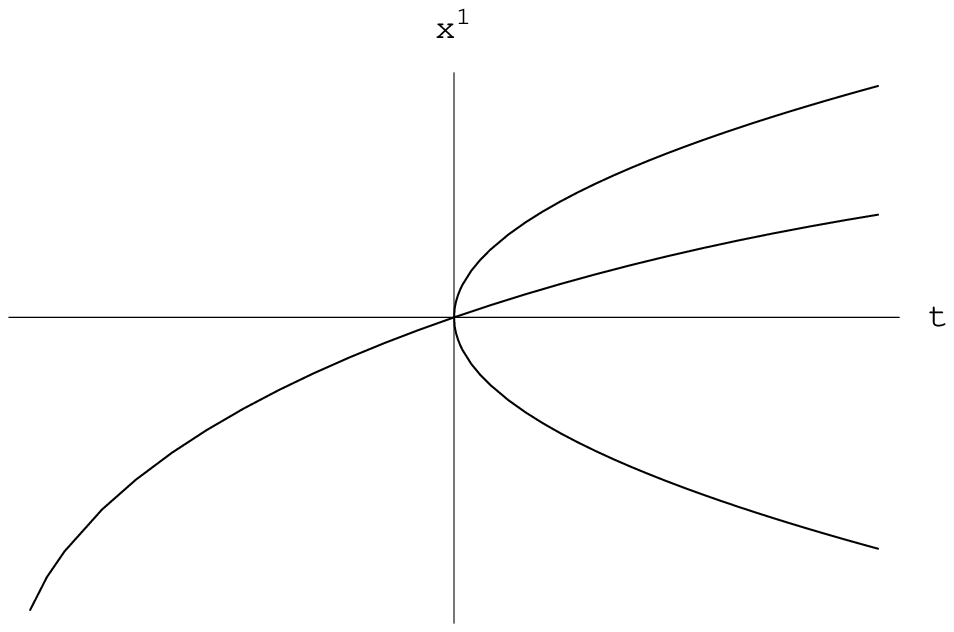,height=5cm,width=5cm}\\
(b)\psfig{figure=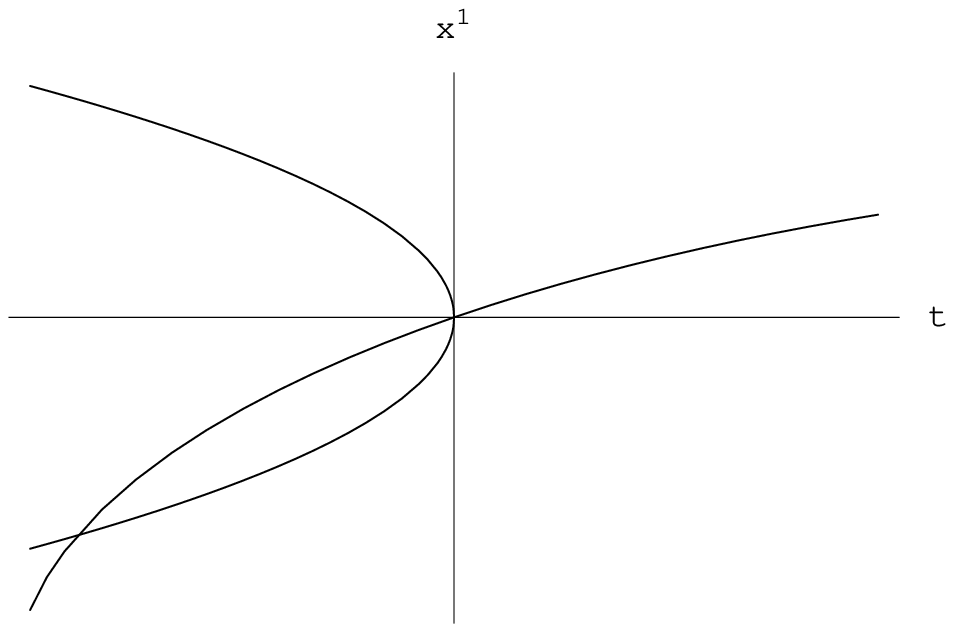,height=5cm,width=5cm}
\end{tabular}
\end{center}
\caption{We fixed the bifurcation point $(x^{1*},t^*)$ at the origin
of $(x^1-t)$ plane. (a) One vortex filament splits into three vortex
filaments at the bifurcation point. (b) Three vortex filaments merge
into one vortex filament at the bifurcation point.} \label{fig4}
\end{figure}

\emph{Case 4 ($A=C=0$)}. Equation(\ref{b1}) and Eq(\ref{b2}) give
respectively
\begin{equation}
\frac{dx^1}{dt}=0,~~~\frac{dt}{dx^1}=0.\label{case4}
\end{equation}
This case is obvious similar to Case 3, see Fig.5.

\begin{figure}[h]
\begin{center}
\begin{tabular}{p{5cm}p{5cm}}
(a)\psfig{figure=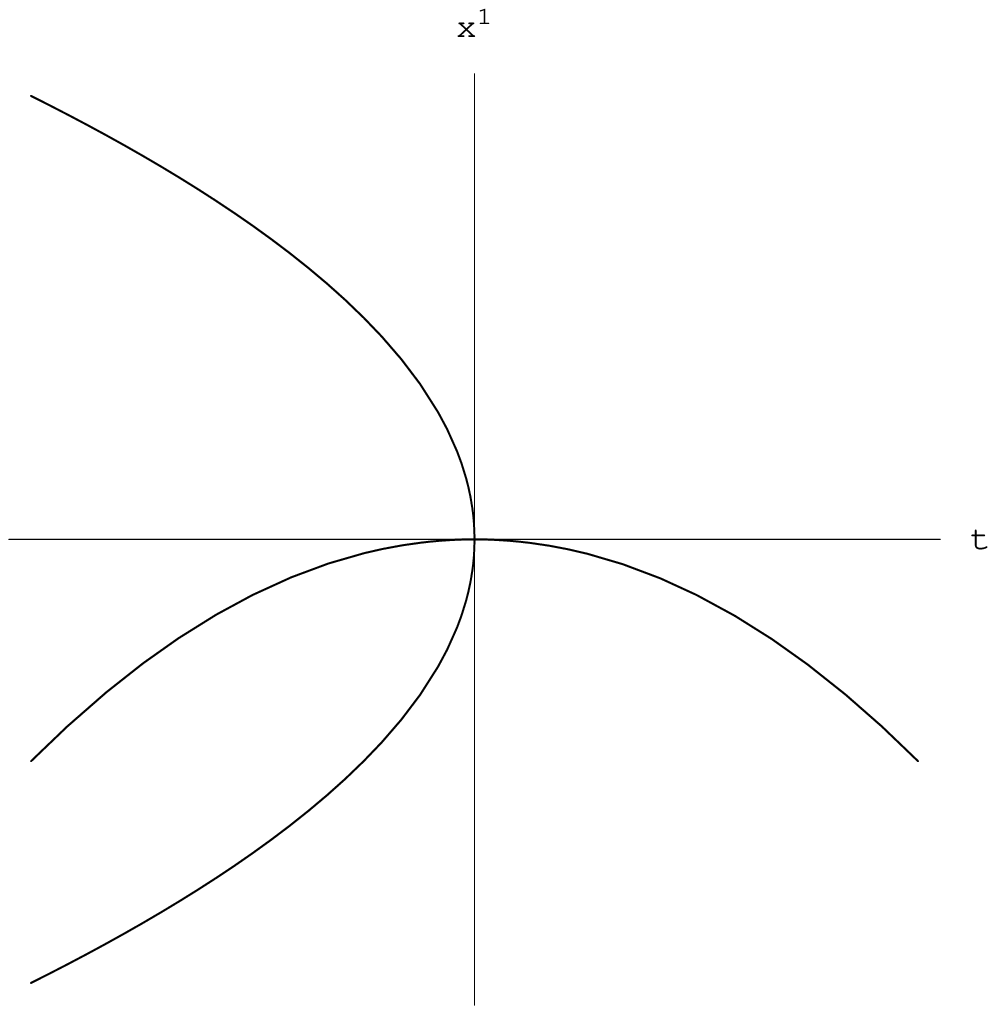,height=5cm,width=5cm}\\
(b)\psfig{figure=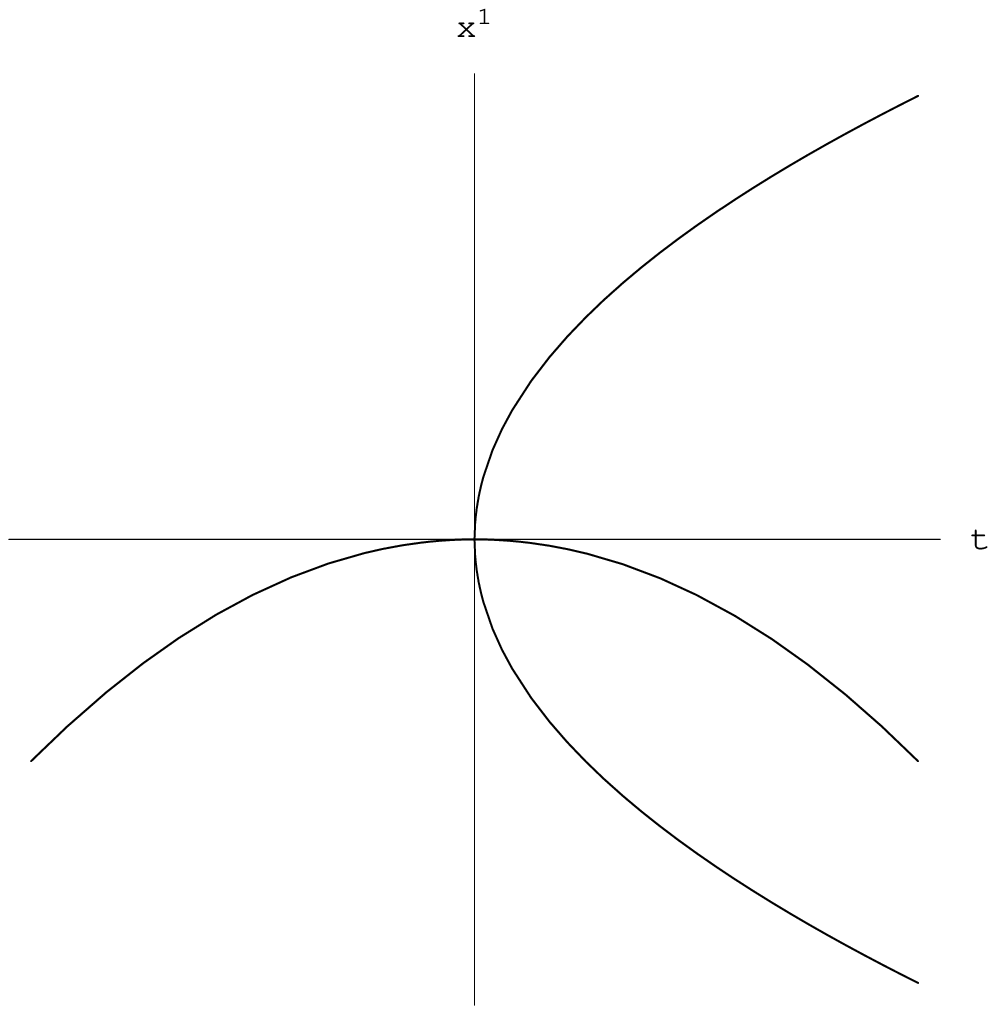,height=5cm,width=5cm}
\end{tabular}
\end{center}
\caption{We fixed the bifurcation point $(x^{1*},t^*)$ at the origin
of $(x^1-t)$ plane. In this case, Two vortex filaments intersect
normally at the bifurcation point. (a) Three vortex filaments merge
into one vortex filament at the bifurcation point. (b) One vortex
filament splits into three vortex filaments at the bifurcation
point.} \label{fig5}
\end{figure}

The above solutions reveal the evolution of the vortex filaments.
Besides the encountering of the vortex filaments, i.e., a vortex
filament pair encounter and then depart at the bifurcation point
along different branch cures [See Fig.2 and Fig.3(a)], it also
includes splitting and merging of vortex filaments. When a
multi-charged vortex filament moves through the bifurcation point,
it may split into several vortex filaments along different branch
curves [See Fig.3(c),  Fig.4(a) and Fig.5(b)]. On the contrary,
several vortex filaments can merge into a vortex filament at the
bifurcation point [See Fig.3(b) and Fig.4(b)].

At the neighborhood of the bifurcation point, we denote scale length
$l=\Delta x$. From Eqs.(\ref{case1})-(\ref{case3}) we can then
obtain the approximation asymptotic relation
\begin{equation}
l\propto (t-t^*).
\end{equation}
The growth rate $\gamma$ or annihilation rate of vortex filament
$\gamma$ of the vortex filament is
\begin{equation}
\gamma\propto const.
\end{equation}
From Eq.(\ref{case4}), one can obtain
\begin{equation}
l=const,~~~~\gamma=0.
\end{equation}
It is obvious that the vortex filaments are relatively at rest when
$l=const$.

The identical conversation of the topological charge shows the sum
of the topological charge of these final vortex filaments must be
equal to that of the original vortex filaments at the bifurcation
point, i.e.,
\begin{equation}
\sum_{i}\beta_{l_{i}}\eta_{l_{i}}=\sum_{f}\beta_{l_{f}}\eta_{l_{f}}\label{cc2}
\end{equation}
for fixed $l$. Furthermore, from the above studies, we see that the
generation, annihilation, and bifurcation of vortex filaments are
not gradually changed, but suddenly changed at the critical points.

\section{Hopf Invariant Constraint On Scroll Wave}
In this section, we will research the topological properties of the
knotted vortex filaments.  We first consider the continuity equation
constraint, from Eq.(\ref{topcu}) one can obtain
\begin{equation}
\partial_i j^i=0,\label{cq3}
\end{equation}
which can also be derived from Eq.(\ref{cq2}). The continuity
equation (\ref{cq3}) implies that the vortex filament may be either
closed loops or infinite curves. In Ref.\cite{topo2}, zhang et.al.
pointed that the continuity equation (\ref{cq3}) is consistent with
the topological rule which governs the scroll wave pinning to an
inclusion\cite{topo3}.

Except the continuity equation constraint on scroll wave, the
complex scroll wave topology may provide other topological
requirements on scroll wave. In the following discussions in this
section, we will study an important knotted invariant which
constraints on scroll wave. It is well know that the Hopf invariant
is an important topological invariant to describe the topological
characteristics of the knot family. In our topological theory of
knotted vortex filaments, the Hopf invariant relates to the
topological characteristic numbers of the knotted vortex filaments
family. In a closed three-manifold $M$ the Hopf invariant is defined
as\cite{hopfscroll,hopf}
\begin{equation}
H=\frac{1}{2\pi}\int_{M}A\wedge
F=\frac{1}{2\pi}\int_{M}A_ij^id^3x.\label{12}
\end{equation}
Substituting Eq.(\ref{10}) into Eq.(\ref{12}), one can obtain
\begin{equation}
H=\frac{1}{2\pi}\sum_{l=1}^{N}W_l \int_{L_l}A_i dx^i\label{helint1}.
\end{equation}
It can be seen that when these $N$ vortex filaments are $N$ closed
curves, i.e., a family of $N$ knots $\xi_l
(l=1,2,\cdot\cdot\cdot,N)$, Eq.(\ref{helint1}) leads to
\begin{equation}
H=\frac{1}{2\pi}\sum_{l=1}^{N}W_l \oint_{\xi_l}A_i dx^i.\label{15}
\end{equation}
This is a very important expression. Consider a transformation of
complex function $Z^{'}=e^{i\theta}Z$, this gives the U(1) gauge
transformation of $A_i: A_i^{'} = A_i + \partial_i\theta $, where
$\theta\in R$ is a phase factor denoting the U(1) gauge
transformation. It is seen that the $\partial_i\theta$ term in
Eq.(\ref{15}) contributes nothing to the integral $H$ when the
vortex filaments are closed, hence the expression (\ref{15}) is
invariant under the U(1) gauge transformation. As pointed out in
Ref.\cite{topo2} , a singular vortex filament is either closed ring
or infinite curve, therefore we conclude that the Hopf invariant is
a spontaneous topological invariant for the vortex filaments in
excitable media.

According to our previous work in Ref.\cite{hopfscroll}, a precise
expression of the Hopf invariant is
\begin{eqnarray}
H=\sum_{k=1}^N W_k^2 SL(\xi_k)+\sum_{k, l=1(k\neq
l)}^NW_kW_lLk(\xi_k, \xi_l),\label{helr}
\end{eqnarray}
where $Lk(\xi_k, \xi_l)$ is the Gauss linking number between
different knotted vortex filaments $\xi_k$ and $\xi_l$, and
$SL(\xi_k)$ is the self-linking number of closed filament $\xi_k$
with an imaginary closed filament infinitesimally
nearby\cite{wri,wri1,gau}. The Eq.(\ref{helr}) reveals the
relationship between $H$ and the self-linking and the linking
numbers of the vortex filaments knots family. Since the self-linking
and the linking numbers are both the invariant characteristic
numbers of the vortex filaments knots family in topology, $H$ is an
important topological invariant required to describe the linked
vortex filaments in excitable media.

In the following we will discuss the conservation of the Hopf
invariant in the branch processes of knotted filaments. In the
branch process of vortex filament, we note that the sum of the
topological charges of final vortex filaments must be equal to that
of the initial vortex filaments at the bifurcation point. This
conclusion is always valid because it is in topological level. So we
have,

(a) for the case that one filament $\xi$ split into two filaments
$\xi_{1}$ and $\xi_{2}$, we have $W_{\xi}=W_{\xi_1}+W_{\xi_2}$;

(b) two vortex filaments $\xi_{1}$ and $\xi_{2}$ merge into one
filaments: $W_{\xi_1}+W_{\xi_2}=W_{\xi}$;

(c) two vortex filaments $\xi_{1}$ and $\xi_{2}$ meet, then depart
as other two filaments $\xi_{3}$ and $\xi_{4}$:
$W_{\xi_1}+W_{\xi_2}=W_{\xi_3}+W_{\xi_4}$.

In the following we will show that when the branch processes of
knotted vortex filaments occur as above, the Hopf invariant is
preserved:

\emph{(A) The splitting case.} We consider one knot $\xi$ split into
two knots $\xi_{1}$ and $\xi_{2}$ which are of the same seif-linking
number as $\xi$ $(SL(\xi)=SL(\xi_{1})=SL(\xi_{2}))$. And then we
will compare the two number $H_\xi$ and $H_{\xi_{1}+\xi_{2}}$ (where
$H_{\xi}$ is the contribution of $\xi$ to $H$ before splitting, and
$H_{\xi_1+\xi_{2}}$ is the total contribution of $\xi_{1}$ and
$\xi_2$ to $H$ after splitting. First, from the above text we have
$W_{\xi}=W_{\xi_{1}}+W_{\xi_{2}}$ in the splitting process. Second,
on the one hand, noticing that in the neighborhood of bifurcation
point, $\xi_{1}$ and $\xi_{2}$ are infinitesimally displace from
each other; on the other hand, for a knot $\xi$ its self-linking
number $SL(\xi)$ is defined as
\begin{eqnarray}
SL(\xi)=Lk(\xi,\xi_{V}),
\end{eqnarray}
where $\xi_{V}$ is another knot obtained by infinitesimally
displacing $\xi$ in the normal direction $\vec{V}$\cite{witten}.
Therefore
\begin{eqnarray}
SL(\xi)=SL(\xi_{1})=SL(\xi_{2})=Lk(\xi_{1},\xi_{2})=Lk(\xi_{2},\xi_{1}),
\end{eqnarray}
and
\begin{eqnarray}
Lk(\xi,\xi'_{k})=Lk(\xi_{1},\xi'_{k})=Lk(\xi_{2},\xi'_{k})
\end{eqnarray}
(where $\xi'_{k}$ denotes another arbitrary knot in the
family($\xi'_{k}\neq \xi, \xi'_{k}\neq \xi_{1,2}$)). Then, third, we
can compare $H_{\xi}$ and $H_{\xi_{1}}+_{\xi_{2}}$ before splitting,
\begin{eqnarray}
H_{\xi}=W^2_{\xi}SL(\xi)+\sum_{k=l(\xi'_{k}\neq
\xi)}^{N}2W_{\xi}W_{\xi'_{k}}Lk(\xi,\xi'_{k}),\label{hopf1}
\end{eqnarray}
where $Lk(\xi,\xi'_{k})=Lk(\xi'_{k},\xi)$; after splitting,
\begin{eqnarray}
H_{\xi_1+\xi_2}&=&W^{2}_{\xi_1}SL(\xi_1)+W^{2}_{\xi_{2}}SL(\xi_{2})+
2W_{\xi_{1}}W_{\xi_{2}}Lk(\xi_{1},\xi_{2})\nonumber\\
&+&\sum_{k=l(\xi'_k\neq
\xi_{1,2})}^{N}2W_{\xi_{1}}W_{\xi'_{k}}Lk(\xi_{1},\xi'_{k})\nonumber\\
&+&\sum_{k=l(\xi'_{k}\neq\xi_{1,2})}^{N}2W_{\xi_{2}}W_{\xi'_{k}}Lk(\xi_{2},\xi'_{k}).\label{hopf2}
\end{eqnarray}
Comparing Eqs.(\ref{hopf1}) and (\ref{hopf2}), we have
\begin{eqnarray}
H_\xi=H_{\xi_1+\xi_2}
\end{eqnarray}
This means that in the splitting process the Hopf invariant is
conserved.

\emph{(B) The mergence case.} We consider two knots $\xi_{1}$ and
$\xi_{2}$, which are of the same self-linking number, merge into one
knot $\xi$ which is of the same self-linking number as $\xi_{1}$ and
$\xi_{2}$. This is obviously the inverse process of the above
splitting case, therefore we have
\begin{eqnarray}
H_{\xi_1+\xi_2}=H_\xi.
\end{eqnarray}

\emph{(C) The intersection case.} This  case is related to the
collision of two knots. we consider that two knots $\xi_{1}$ and
$\xi_{2}$, which are of the same self-linking number, meet, and then
depart as other two knots $\xi_{3}$ and $\xi_{4}$ which are of the
same self-linking number as $\xi_{1}$ and $\xi_{2}$. This process
can be identified to two sub-processes: $\xi_{1}$ and $\xi_{2}$
merge into one knot $\xi$, and then $\xi$ split into $\xi_{3}$ and
$\xi_{4}$. Therefore, from the above two cases (B) and (A) we have
\begin{eqnarray}
H_{\xi_1+\xi_2}=H_{\xi_3+\xi_4}
\end{eqnarray}
Therefore we acquire the result that, in the branch processes during
the evolution of knotted vortex filaments (splitting, mergence, and
intersection), the Hopf invariant is preserved.

The above analysis show that the branch processes of knotted vortex
filament family must satisfy the Hopf invariant constraint. This
conclusion is obtained only from the viewpoint of topology without
using any particular models or hypothesis. Therefore, the Hopf
invariant is a more extensive topological constraint on scroll wave,
and it is valid in almost systems which support the existence of
scroll wave.

According to Winfree and Stogatz\cite{knot3}, there is an
``exclusion principle" governed the scroll wave knotting and linking
through each other in chemical system. The chemical requirement
plays a crucial role in such ``exclusion principle". It states that
the topology of scroll wave in such system must satisfy the
constraints that:
\begin{eqnarray}
W_k SL(\xi_k)+\sum_{l=1 (l\neq k)}^{N}W_l Lk(\xi_k,\xi_l)=0.
\end{eqnarray}
It is very easy to see that the ``exclusion principle" makes the
Hopf invariant trivially, i.e., $H=0$. This is just a special case
of the Hopf invariant constraint. When branch processes of scroll
wave occur, it is obvious from above discussion that the ``exclusion
principle" is also protected by topology.
\section{Conclusion and Discussions}
First, we give a prime review of the topological theory of vortex
filaments in three dimensional excitable media. When $D(\phi/x)=0$,
the intersection, splitting, and merging of line defects in
three-dimensional space are investigated in detail by making using
of Duan's topological current theory. Second, the evolution of
vortex filaments in ($3+1$)-dimensional space-time is studied. There
exist crucial cases of branch processes in the evolution of vortex
filaments when the Jacobian $D(\phi/x)=0$, i.e., $\eta_l$ is
indefinite. At one of the limit points of the complex function $Z$,
a pair of vortex filaments with opposite topological charge can be
annihilated or generated. At one of the bifurcation points of $Z$, a
vortex filament with topological charge $W_\xi$ may split into
several vortex filaments (total topological charges is $W_\xi$);
conversely, several vortex filaments (total topological charges is
$W_\xi$) can merge into one vortex filament with a topological
charge $W_\xi$. Also, at one of the bifurcation points of $Z$, two
filaments meet and then depart. These show that vortex filaments are
unstable at these branch points of $Z$. From the topological
properties of the complex function $Z$, we obtained that the
velocity of the vortex filaments is infinite when they are being
annihilated or generated, which agrees with the similar results of
line defects what was obtained by Bray and Mazenko\cite{bra}. The
velocity of the vortex filament at the limit point or bifurcation
point has been shown clearly in Fig.1 to Fig.5 (The slope of the
curve). Third, based on the branch process of vortex filaments, it
is showed detailed that the Hopf invariant of knotted scroll wave is
preserved in the branch process, this is consistent with our
proposal that the Hopf invariant may implies a new constraint (the
Hopf invariant constraint) on scroll wave. Furthermore, it also
revealed that the ``exclusion principle" in some chemical media is
just the special case of the Hopf invariant constraint, and during
the branch processes the ``exclusion principle" is also protected by
topology. Finally, we would like to point out that both branch
theory of vortex filaments and the Hopf invariant constraint in this
paper are obtained from only the viewpoint of topology, without
using any particular models or hypotheses. These results are valid
for all systems which support the existence of scroll wave.

In this paper, we give a rigorous and general topological
investigation of scroll wave. This work can be applied in practice.
Here we give some discussions about how to connect our work to the
phenomenology and mathematical analysis of scroll wave. \emph{(i).}
The regular condition (\ref{recon}) can be regarded as stability
condition of scroll wave. When the solutions of Eq.(\ref{rd}) is
determined, the exact expression of (\ref{recon}) can be work out.
In this case, we can directly calculate the stability condition of
scroll wave. This provide a direct approach to investigate the
stability of scroll wave and how to control it. \emph{(ii).} When
the regular condition (\ref{recon}) fails, the scroll wave is
unstable. In this case, the branch processes occur. The branch
condition $D(\phi/x)=0$ predicts where and how these processes will
occur, if the scroll wave solution of Eq.(\ref{rd}) is work out. In
laboratory, these branch processes can be produced by using the
branch condition. This provide an experimental approach to test our
work. \emph{(iii)} In this paper, we have predicted that the
velocity of the vortex filaments is infinite when they are being
annihilated or generated. The similar phenomena has been obtained in
the phase-ordering system\cite{bra}. We expect that this phenomena
will be observed for scroll wave in laboratory. \emph{(iv).} For
knotted vortex filaments, the Hopf invariant will play an important
role in control the behaviours of scroll wave. During the branch
processes, the Hopf invariant will protect by topology. We also hope
that this will be proved by experiment. \emph{(v).} At last, we
point out that Eq.(\ref{rd}) determines the solution of $Z$, and of
course it determines the regular condition and branch condition. The
investigation presents in this paper is based on an important
precondition that the scroll wave solutions can be worked out from
Eq.(\ref{rd}).

\begin{acknowledgments}

This work was supported by the National Natural Science Foundation
of China and the Cuiying Programm of Lanzhou University, P. R.
China.
 \end{acknowledgments}

 \end{document}